**Title:** Robust Super-Moiré in Large Angle Single-Twist Bilayers

Yanxing Li[†,1], Chuqiao Shi[†,2], Fan Zhang[†,1], Xiaohui Liu[†,1], Yuan Xue[1], Viet-Anh Ha[1,4], Qiang Gao[3], Chengye Dong[5], Yu-chuan Lin[5,6], Luke N Holtzman[7], Nicolas Morales-Durán[1], Hyunsue Kim[1], Yi Jiang[8], Madisen Holbrook[9], James Hone[9], Katayun Barmak[7], Joshua Robinson[5], Xiaoqin Li[1], Feliciano Giustino[1,4], Eslam Khalaf[3], Yimo Han (yh76@rice.edu)[2], Chih-Kang Shih (shih@physics.utexas.edu)[1]

[1] Department of Physics, The University of Texas at Austin, Austin, TX, USA.

[2] Department of Materials Science and NanoEngineering, Rice University, Houston, TX, USA.s

[3] Department of Physics, Harvard University, Cambridge, MA, USA.

[4] Oden Institute for Computational Engineering and Sciences, The University of Texas at Austin, Austin, TX, USA.

[5] Department of Materials Science and Engineering, Pennsylvania State University, University Park, PA, USA

[6] Department of Materials Science and Engineering, National Yang Ming Chiao Tung University, Hsinchu, Taiwan.

[7] Department of Applied Physics and Applied Mathematics, Columbia University, New York City, NY, USA

[8] Advanced Photon Source, Argonne National Laboratory, Lemont, IL, USA.

[9] Department of Mechanical Engineering, Columbia University, New York City, NY, USA

† These authors contributed equally: Yanxing Li, Chuqiao Shi, Fan Zhang, and Xiaohui Liu.

**Abstract:** Forming long wavelength moiré superlattices (MSL) at small-angle twist van der Waals (vdW) bilayers has been a key approach to creating moiré flat bands[1–8]. The small-angle twist, however, leads to strong lattice reconstruction, causing domain walls and moiré disorders, which pose considerable challenges in engineering such platforms[6–10]. At large twist angles, the rigid lattices render a more robust, but shorter wavelength MSL, making it difficult to engineer flat bands. Here, we depict a novel approach to tailoring robust super-moiré (SM) structures that combines the advantages of both small-twist and large-twist transition metal dichalcogenides (TMDs) bilayers using only a single twist angle near a commensurate angle. Structurally, we unveil the spontaneous formation of a periodic arrangement of three inequivalent commensurate moiré (CM) stacking, where the angle deviation from the commensurate angle can tune the periodicity. Electronically, we reveal a large set of van Hove singularities (VHSs) that indicate strong band hybridization, leading to flat bands near the valence band maximum. Our study paves the way for a new platform of robust SM bilayers with structural rigidity and controllable wavelength, extending the investigation of the interplay among band topology, quantum geometry, and moiré superconductivity to the large twist angle regime.

**Main Text:**

The ability to tailor 2D electronic superlattice by forming moiré patterns in vdW bilayers at a small angle has attracted intense research activity over the past few years. More recently, the moiré designs have expanded to include SM patterns in multi-twist multi-layers arising from the interference of different moiré patterns[7,11–14]. The small-angle twist results in a long wavelength moiré periodicity which is advantageous for the formation of flat bands[1]. However, when the twist angle is small, moiré disorders occur due to strong atomic lattice reconstruction, leading to structural heterogeneity within and across samples[8]. In addition, the lattice reconstruction in small-twist moiré also causes domain wall formation, whose structure dominates the underlying physics[15], making it difficult to probe intrinsic correlation under SM. Large-angle moiré structures, while being structurally more rigid, have shorter wavelengths and usually lack the strong hybridization of the bands near the fermi energy that leads to strong correlation physics[16–19]. Here, we show a novel approach to the formation of a SM structure in TMDs (i.e. WSe$_2$ in our case) bilayers twisted near the commensurate angle of 32.2° that simultaneously combines many desirable properties of existing moiré and SM structures: (i) it is associated with a length scale of the same order as small twist angle moiré systems, resulting in narrow electronic bands, (ii) it features a very smooth moiré pattern where the effects of lattice reconstruction are negligible, and (iii) it is achieved by tuning a *single twist angle* in a bilayer with flat band engineerability.

The scanning transmission electron microscopy (STEM) and scanning tunneling microscopy (STM) investigations reveal that the SM structure consists of a periodic arrangement of three distinctly different commensurate stacking arrangements, called commensurate rotational faults (CRF), relating to each other by a small gliding vector (in order of 0.5 Å). The small angle deviation, δ, from the exact commensurate angle (32.2°), endows a spatial dependence of the gliding vectors, thus resulting in a periodic arrangement of three distinct CRFs with a tunable SM wavelength of $\lambda_{SM} \propto \delta^{-1}$. Notably, the SM structure leads to the emergence of narrow bands, identified through quasiparticle interference (QPI) and STS with a large set of VHSs, which are also confirmed using theoretical calculations. Our discovery demonstrates the engineering of electronic band flatness at robust large-angle twisted bilayers, offering a powerful new platform to tailor novel electronic structures in twistronics.

**CRFs and SM**

The CM occurs at discrete angles that satisfy the condition of $\theta_t = \arccos(\frac{p^2+4pq+q^2}{2(p^2+pq+q^2)})$, where p and q are co-prime integers. At twist angle $\theta_t = 27.8/32.2°$, it leads to a CM structure with a $\sqrt{13} \times \sqrt{13}$ unit cell. Noted that for CM, there exist different types of CRFs[19]: sublattice exchange even (SE$_{even}$) versus sublattice exchange odd (SE$_{odd}$) relating to each other by a lateral gliding vector. For TMDs $\theta_t = 32.2°$ CM, there are two different SE$_{odd}$ structures, one centering around the transition metal atoms (SE$_{odd\_W}$), and one around the chalcogen atoms (SE$_{odd\_Se}$)[20].

We first discuss the structural aspects of $\theta_t = 32.2°$ CM. STEM (Fig. 1a) is ideal for visualizing the atomic structure of multilayer vdW materials due to the scattering by atomic nucleus from all layers and its sub-angstrom spatial resolution[21,22]. Fig. 1b depicts three distinct types of CRFs at the commensurate angle of 32.2° based on their initial non-twisted stacking orders, rotational axis, and symmetry (Supplementary Figure S1). The simulated annular dark field (ADF) images (left) and experimental ADF-STEM images (right) of SE$_{even}$ (top), SE$_{odd\_Se}$ (middle), and SE$_{odd\_W}$ (bottom) patterns are shown in Fig. 1c, respectively. It should be noted that observations of three distinct CRFs have also been reported previously in TaS$_2$ and MoS$_2$ bilayers[20]. While perfect 32.2° leads to a single CRF, by applying a small extra twist (δ) from the commensurate angle ($\theta_{CM}$), i.e. $\theta_t = \theta_{CM} \pm \delta$, one finds a periodic arrangement of three CRFs, creating a SM structure with a wavelength of $\lambda_{SM} \approx \frac{\sqrt{3}d_1 \sin(\theta_{CM})}{2\sin(\delta/2)}$ (see Fig. 1d), where $d_1$ is the SM gliding distance between the same CRFs, which will be discussed later. Shown in Fig. 1e is the example for $\theta_t = \sim 32.2°$ (deviation within 0.2°), where we observe long-range order of CRFs due to the large wavelength of SM structure. The precise angle is measured by the nanobeam four-dimensional STEM (4D-STEM) (Supplementary Figure S2). On the other hand, in Figure 1f with $\theta_t = 31.5°$, a SM pattern with $\lambda_{SM} \approx 8$ nm and three CRFs emerge, as expected from the equation of $\lambda_{SM}$ (4D-STEM analysis of this sample is shown in Supplementary Figure S3). This contrasting behavior is also reflected in the FFT images (Fig. 1g). For $\theta_t = 31.5°$ sample, the intrinsic CM frequency peaks at $q = q_M$ are split into 3 dots due to the formation of SM, while the $\theta_t = 32.2°$ sample remains single dots. By reconstructing the real space Fourier amplitude image[23] using the CM frequency peaks at $q = q_M$ marked by red diamonds in Fig. 1f, the long-range fluctuated CRFs and SM pattern with periodically arranged CRFs are depicted respectively in Fig. 1h (see detailed analysis in Supplementary Figure S4). Note that the observed long-range CRF fluctuation in samples with very small deviation angles (< 0.2°) is likely attributable to sample preparation and suspension over a hole in the TEM grid. Additional experiments involving the encapsulation of the twisted bilayer in hBN demonstrated more uniform individual CRF patterns near 32.2°, suggesting that transferring the sample onto a flat substrate can effectively mitigate such fluctuations.

**Gliding translation relation**

The geometric relation between the three CRFs[20] is central to understanding the formation of SM structures. Figure 2a shows the gliding translation diamond among the CRFs for 32.2° CM, where each CRF unit cell can be transited to another by a gliding translation of $d_2 = 52$ pm. One CRF unit cell can also glide by a displacement as small as $d_1 = 90$ pm to return to itself. The 31.5° sample is further analyzed by electron ptychography technique enabled by large-convergence angle 4D-STEM, which improves the spatial resolution to deep-sub-angstrom level[21,22]. Within each SM unit cell, we can capture the gradual gliding transition between each pair of these CRFs. Specifically, in Figure 2b (also in Supplementary Figure S5), the transitions from $SE_{even}$ to $SE_{odd\_Se}$, $SE_{odd\_Se}$ to $SE_{odd\_W}$, and $SE_{odd\_W}$ to $SE_{even}$ are displayed. Such transitions establish a closed loop, which rationalizes the translational symmetry in the SM. The schematic of these gliding displacement vectors in real space lattice is shown in Figure 2c where the minimum displacement vectors needed from the initial $SE_{odd\_Se}$ pattern to $SE_{even}$, $SE_{odd\_W}$, and itself are depicted respectively. For each transition between any two of the CRFs, there are three angular equally spaced directions (see details in Supplementary Figure S6). As a result, the formation of a 3-fold SM structure is guaranteed. We demonstrate such 3-fold rotational symmetry by the magnified ADF image in Figure 2d, where the center of each hexagonal tile is $SE_{even}$ and the $SE_{odd}$ patterns sit at the six corners. Among the six corners, half are $SE_{odd\_Se}$ and half are $SE_{odd\_W}$, forming three sublattices in the SM structure, each retaining a 3-fold symmetry.

To confirm the commensurate nature of the SM pattern, we simulated a perfect SM structure with periodic CRF by introducing a periodic lattice distortion (PLD) model[24]. The PLD model entails a large-angle moiré pattern with a periodic translational distortion to each layer with opposite chirality (see details in Supplementary Note 1). Although minor twist-angle deviations can introduce discrepancies in each CRF, these were effectively corrected in the simulation by applying localized adjustments through the PLD. Comparison of diffraction patterns with and without the PLD revealed only minor differences in the intensity of higher-order spots (Supplementary Figure S8-9). The model further predicted atomic displacements on the order of a few picometers, which is an order of magnitude smaller than small-angle lattice distortion described by the same model[20,24,25]. This is also beyond the resolution of ADF or electron ptychography imaging, and thus no obvious difference was observed experimentally. These results demonstrate that the experimentally observed SM structure aligns closely with the simulated perfect SM, confirming that large-angle SM patterns mitigate lattice relaxation issues.

**Narrow bands in the SM lattices**

The SM structures that consist of three different CRFs are anticipated to host novel electronic properties. These are investigated using STM/S (Supplementary Figure 10). Shown in Fig. 3a is a 20 nm field of view (FOV) STM image for a ~31.1° sample. As discussed previously[16,26], in STM imaging, the CM and SM structure at a large twist angle can be revealed only if the bias is set at the semiconducting gap, namely, the tunneling occurs between the tip and the underlying graphene[26,27]. The mechanism for image contrast is further discussed in Supplementary Note 2. The CM unit cell is labeled in Fig. 3a. In this image, the SM structure is difficult to discern due to the interference of large corrugation associated with the atomic features and the influence of the underlying graphene substrate (details discussed in Supplementary Note 3). However, the fast FFT image in Fig. 3b reveals clear splitting in each order of the 32.2° CM frequency peaks, marked by blue circles and red diamonds. To elucidate the SM pattern, we select Fourier signals within the blue circles and reconstruct the real-space image shown in Fig. 3c that exhibits a distinct SM pattern with a periodicity ($\lambda_{SM}$) of approximately 4 nm. Similarly, longer wavelength SM is observed in a ~31.5° twisted WSe$_2$ sample whose FFT image also reveals the splitting feature (Extended Figure 1).

STS is conducted along the direction of the SM unit cell (which is also the diagonal direction of the CM unit cell), as labeled by sites in Fig. 3a. Fig. 3d shows the spectra acquired by the conventional STS. The two prominent peaks in the VB correspond to the split Γ valleys resulting from the interlayer hybridization. The energy splitting of Γ valleys is $\Delta_{\Gamma_1-\Gamma_2}$ = 610 meV, which is similar to that observed in the $\theta_t = 30°$ quasicrystal[16]. Note that the conventional STS usually does not reveal spectral features associated with the K-valley due to the large decay constant into the vacuum. This difficulty can be mitigated by performing the constant current scanning tunneling spectroscopy (CCSTS) where the tip-sample distance is automatically reduced when the tunneling occurs in the K-valley[28]. The behavior of tip-sample distance in CCSTS measurement (Z-V) is shown in Supplementary Figure S14. In CCSTS, the spectral features are revealed by the differential conductance, $(\partial I/\partial V)_I$, shown in Fig. 3e. Most strikingly, at least eight prominent peaks belong to the K-valley states. Among them, the three peaks close to the K valence band maximum ($K_{VBM}$) are inherently strong, probably indicating the logarithmic divergent VHSs. To aid the interpretation, we first discuss the intrinsic electronic structures of $\theta_t = 32.2°$ CM. The inset of Fig. 3f shows the Brillion zones (BZs) for the top layer (red), bottom layer (dark blue), and CM (light blue), respectively. $\theta_t = 32.2°$ CM is expected to host only one set of K-valley anti-crossing above the $\Gamma_1$ which takes place at the $m$ point of CM BZ. This anti-crossing occurs due to the Umklapp scatterings in its third and fourth order (See details in Supplementary Note 4). In this case, only two VHSs are expected due to the moiré mini-gap at the $m$ point, significantly different from what we observe here in the $\theta_t \sim 31.1°$ sample.

One might argue that these extra peaks are simply phonon sidebands resulting from inelastic scatterings. Note that electron-phonon scatterings have rather low cross-sections and are typically observable only in second derivatives, $d^2I/dV^2$, in tunneling spectra[29–31]. Moreover, such inelastic tunneling channels involving phonons occur with equal likelihood for either tunneling into or out of the sample (particle-hole symmetry). In our case, only the valence band K-valleys exhibit multiple sharp peaks that are absent in the conduction band (Supplementary Figure S14). As we discuss below, these spectral features as well as the interference patterns are due to SM scatterings, resulting in the zone folding of the original CM band structures.

**SM scatterings**

Shown in Fig. 4a are the SM BZs (grey), the CM BZ (purple), the top, and bottom atomic layer BZs (red and blue) for the SM structure formed at $\theta_t = 31.1°$, respectively. At this angle, $\lambda_{SM} \sim 2\sqrt{3}\ \lambda_{CM}$, and the SM is also commensurate with the CM, reflected in Fig. 4a. The SM BZs exhibit a unique $n\sqrt{3} \times n\sqrt{3}$ superlattice reconstruction within the CM BZs (n is an integer. For $\theta_t = 31.1°$, n = 2). The relative orientation of SM BZs and CM BZs enables the K valleys of individual layers to overlap with the SM Γ valleys, rearranging the K valleys as a triangular lattice. To capture the electronic structure, we apply a continuum model[1,32,33] described in Supplementary Note 5. We only consider the K-valley states since the Γ-valley states are ~ 0.3 eV below the VBM. There are three key parameters in the model Hamiltonians, the interlayer hopping term ω, the SM scattering term $V_{SM}$, and the geometric phase term ψ. ω is the term that endows the mini-gap in the CM shown in Fig. 3f. One can estimate ω to be ~ 30 meV, by adopting the mini-gap value being observed in the moiré quasicrystal involving a similar order of Umklapp processes[16]. The SM scattering term, $V_{SM}$, is treated as an adjustable parameter. The resulting SM potential profile (after incorporating the geometric phase ψ) is plotted in Fig. 4b. Shown in Fig. 4c are electronic structures for the SM structure at $\theta_t = 31.1°$, calculated using ω = 30 meV and $V_{SM}$ = 15 meV, with the density of states (DOS) (blue curve) shown in Fig. 4e. Also shown in Fig. 4e, is an experimentally measured spectrum (purple). The theoretically calculated DOS captures the main prominent features observed experimentally (See the whole comparison in Supplementary Fig. S18). We have also used DFT calculations to estimate the parameter $V_{SM}$ and obtained a value of 5 meV. The result is shown in Supplementary Figure S20 albeit with slightly less agreement with the experiment. We emphasize, the most prominent minigap observed experimentally occurs at about -0.75 eV with a size of ~ 30 meV. This gap size sets the magnitude of $V_{SM}$.

As expected, when the SM periodicity is changed, the electronic structure changes accordingly. Figure 4d, f, show the theoretical result for SM structure at $\theta_t = 31.4°$ with $\lambda_{SM} \sim 3\sqrt{3}\ \lambda_{CM}$. Two nearly degenerate flat bands emerge near the VBM (Figure 4d, f), a result of the equivalence between the top and bottom layers. This degeneracy can be lifted by applying a vertical field (Supplementary Fig. S21). These flat bands can facilitate a platform to explore strongly correlated phenomena[34,35].

## QPI

With the above band structure pictures from the continuum model, we next analyze the QPI patterns at a wide range of bias voltages (Supplementary Fig. S23) in a relatively clean region (Fig. 5a). Shown in Fig. 5b-c are conductivity images acquired at two different biases in the K-valley, exhibiting wave interference patterns. Instead of the ring-like patterns observed in samples with very dilute defect density (shown in Extended Fig. 2a-c), here, the QPIs show stripe patterns. Note that QPI patterns depend on the density of scattering centers, and their spatial arrangement[36]. The electron structure information is contained in the energy dependence of the wave vector (so-called E vs. q[37]). The challenge in acquiring QPIs for K-valley states is that the tip-to-sample distance is significantly reduced. To keep a long sequence of bias-dependent conductivity images without encountering tip-change, we limit the FOV to 8 nm by 8 nm. Since the stripe patterns have a well-defined direction (along which the line traces are extracted and shown in Fig. 5d), we can perform 1D Fourier analysis as shown in Fig. 5e deduced from the bias range from -0.68 V to -0.88 V. The resulting E versus q/2 is plotted in Fig. 5g along with the calculated band structures. The plot contains two sets of data acquired at different regions with consistent phase alternating phenomena (Supplementary Note 6). The electronic structure is plotted along the $\gamma - m - \gamma$ direction where the SM Bragg plane intersects at m. Here the initial state $|k_i\rangle$ and the final state $|k_f\rangle$ have the same magnitude but opposite directions. Thus one expects a q vector with a magnitude of $|q| = 2|k|$, as illustrated in Fig. 5f. The mapping of E versus q/2 (Fig. 5g) agrees quite well with the calculated band structure along the $\gamma - m - \gamma$ direction. In addition to E vs q/2, we note that the conductivity image acquired in the SM gap (taken at -0.75 V, shown in Fig. 5c) and that acquired below the SM gap (at -0.76 V, Fig. 5b), are roughly out of phase (See the complete phase analysis in Supplementary Fig. S24). This is also consistent with the interpretation that this prominent gap is due to the scattering of SM Bragg plane. Data acquired at another region also shows such a phase reversal for QPI patterns acquired above/below the mini-gap (Supplementary Fig. S25). Note that the QPI data acquired for ~31.5° SM (at 77 K) with low defect density ($<10^{11}cm^{-2}$) also yields a consistent dispersion (Extended Figure 1). The QPI analysis thus provides a consistent picture that the electronic structures observed using STS is a result of zone folding due to SM scatterings.

**Discussion and conclusion**

In summary, by tuning a twist bilayer of $WSe_2$ near the commensurate angle, we show a new platform to engineer SM structures that harbor novel electronic structures. The SM structure results from the symmetry breaking of an otherwise homogenous moiré crystal, leading to a periodic arrangement of three distinct CRF domains, whose periodicity can be tuned via an extra tuning angle near the commensurate angle. The narrow band features are observed here via STS investigations which show a large set of VHSs, manifested by the SM scattering. These spectral features are well captured by the theoretical analysis using a continuum model.

The ability to design the SM structure using only one large twist angle is very attractive. On one hand, the structure is formed with a large twist angle, thus offering the advantage of structural rigidity, free of large-scale lattice reconstruction and domain wall formation typically encountered in the small angle moiré structure (Supplementary Fig. S26). On the other hand, the long wavelength SM structure is controlled by the small angular deviation, similar to that for the small angle moiré, thus retaining the flat-band engineering feature. It therefore combines the advantages of both small and large twist angles. The structural transition between sub-units within the SM is expected to be smooth and leads to smooth scattering potential, an attractive feature for theoretical modeling. The SM scattering results in the formation of narrow bands, which will host strongly correlated phenomena and exotic quantum phases. Such novel SM structures provide an ideal platform for future investigation of the interplay among the band topology[38,39], quantum geometry[40], and unconventional superconductivity[41,42] at large angle twisted TMD bilayers.

**Figure Captions**

**Figure 1 | CM and SM patterns in near 32.2° twisted bilayer WSe$_2$. a,** Schematic for the ADF-STEM and electron ptychography. **b,** The unit cells for the three CRFs of 32.2° CM. Where blue and grey atoms represent the W and Se atoms, respectively. The red open circles mark the center overlapped sites for the atoms from two layers. **c,** Simulated ADF-STEM image for three CRFs of 32.2° CM (left) and experimental ADF-STEM images of the CRFs in the 31.5° sample (right). The colors of picture borders correspond to different CRFs as assigned in b. Orange represents SE$_{even}$ (top), Purple represents SE$_{odd\_Se}$ (middle) and red represents SE$_{odd\_W}$ (bottom). The red circles label the center overlapped sites for the atoms from two layers. **d,** Curve: model and simulation results of twist-angle v.s SM periodicity. Schematics: Tuning the SM near 32.2° commensurate angle. When the angle is precisely 32.2°, only one of the CRFs will exist (center panel, with red representing SE$_{odd\_W}$). When the angle deviates, a SM with different CRFs appears (left and right panels, orange: SE$_{even}$, purple: SE$_{odd\_Se}$, gray: transition region). **e,** ADF-STEM image of a ~32.2° twisted bilayer WSe$_2$. **f,** ADF-STEM image of a ~31.5° twisted bilayer WSe$_2$. The yellow hexagon and the white dashed diamond mark the super moiré tile and the super moiré unit cell respectively. Colored dots mark different CRFs accordingly. **g,** The FFT images of the STEM images in e (left) and f (right). Where the six red diamonds label the zeroth order moiré frequency peaks, the black squares mark the Bragg peaks. **h,** Real-space amplitude image from the inversed FFT (iFFT) of the selected q from the moiré frequency peaks (red diamonds in g). Left and right panels represent the iFFT images for ~32.2° and ~31.5° samples, respectively. Scale bars: c, 1nm. e,f,h, 4 nm.

**Figure 2 | Gliding translation relation and SM symmetry. a,** The gliding translation diamond for describing the gliding relations among the CRFs. d$_1$ and d$_2$ represent the smallest gliding displacement for the transitions between two same CRFs and different CRFs, respectively. **b,** The electron ptychography images from ~31.5° sample. Top: Transition from SE$_{even}$ (left hexagon) to SE$_{odd\_Se}$ (right shield). Middle: Transition from SE$_{odd\_Se}$ (left shield) to SE$_{odd\_W}$ (right shield). Bottom: Transition from SE$_{odd\_W}$ (left shield) to SE$_{even}$ (right hexagon). **c,** The displacement vectors needed for the current SE$_{odd\_Se}$ configurations to transit to the other CRFs. Different colors represent the displacement vectors for different CRFs, according to the colors assigned in a. **d,** The experimental ADF-STEM image of a ~31.5° twisted bilayer WSe$_2$. The yellow shields represent two types of SE$_{odd}$ configurations. The shields pointing down are SE$_{odd\_Se}$ while the one pointing up is SE$_{odd\_W}$). The yellow hexagons depict the SE$_{even}$ cells. The white dashed lines label the boundaries of hexagonal tiles. Scale bars: b, d, 1 nm.

**Figure 3 | Multiple VHSs in the SM lattice. a,** A 20nm FOV STM image of a ~31.1° twisted bilayer WSe$_2$. The dots and dashed line mark the sites of measurements and the direction respectively. The moiré unit cell is exemplified by the dashed diamond. ($V_{Bias}$ = −0.3 V, I = −50 pA) **b,** FFT image of the STM image in a. red diamonds, blue circles, and black squares mark the first order, third order moiré frequency peaks and the Bragg peaks, respectively. **c,** Real-space amplitude image for q = 2q$_M$ by applying inverse FFT to the blue circle masks as labeled in b. **d,** Constant height STS taken at the sites shown in a. The blue arrows mark the two upmost Γ valleys. Spectra are shifted for better visibility. ($V_{set}$ = −2.0 V, I = −50 pA and $V_{amp}$ = 19 mV) **e,** Constant-current ∂I/∂V spectroscopies from the measurement sites 1-4. The energy locations of Γ$_1$ and K$_{VBM}$ are indicated by the blue and red arrows, respectively. Spectra are shifted for better visibility. ($V_{int}$ = −1.2 V, I = −20 pA and $V_{amp}$ = 10 mV) **f,** The simulated DOS spectrum for a 32.2° CM using the 2D free electron gas model under a spin-conserved condition. Inset: schematic showing the first Brillouin zones for 32.2° twisted WSe$_2$ bilayer. The light blue, red, and dark blue hexagons mark the moiré, top-layer, and bottom-layer Brillouin zones, respectively. Scale bars: a, c, 2 nm.

**Figure 4 | Continuum model for the SM lattice. a,** The SM BZs of a 31.1° twisted bilayer WSe$_2$. The top layer, bottom layer, 32.2° CM, and 31.1° SM BZs are, respectively, marked by the red, blue, purple, and grey hexagons. The top layer BZ of the 32.2° twisted bilayer WSe$_2$ before the angle mismatch is depicted by the light red hexagon. **b,** The real space scattering potential landscape fitted with DFT calculated band alignments of a 31.1° twisted bilayer WSe$_2$. Orange, purple, and red dots label the SE$_{even}$, SE$_{odd\_Se}$ and SE$_{odd\_W}$ configurations respectively. The black dashed diamond marks the SM unit cell. **c,** The folded band structure of 31.1° SM simulated by the continuum model. **d,** The folded band structure of 31.4° SM simulated by the continuum model. **e,** DOS spectrum of the band structure in c. The purple line represents the CCSTS taken from site 1, as also shown in Fig. 3e. **f,** DOS spectrum of the band structure in d.

**Figure 5 | QPI in a ~31.1° SM. a,** A topography image near a weak defect taken simultaneously with the constant current dI/dV mapping shown in b at $V_{Bias}$ = −0.76 V, I = −20 pA. **b,** Constant current dI/dV mapping taken at $V_{Bias}$ = −0.76 V, I = −20 pA. **c,** Constant current dI/dV mapping taken at $V_{Bias}$ = −0.75 V, I = −20 pA. **d,** The traces along the wave propagating direction extracted from (the white dashed lines in b, c) constant current dI/dV mappings taken in region 1 (SI Fig. S23) at several bias voltages. **e,** The 1D Fourier Transfer of the traces from d. Black arrows mark the frequency peaks representing the QPI q wavelength. The inner peaks originate from the finite sampling of these traces. **f,** Schematic of the intra-K-valley scattering in 31.1° SM. The purple and grey BZs depict the CM and SM BZs, respectively. $\Delta_{SM}$ is the SM minigap while $\Delta_{CM}$ is the CM minigap. The red arrow exhibits the q vector from intra-K-valley scatterings. The SM γ overlaps with CM $\Gamma_m$ and $K_m$ valleys. **g,** E v.s. q/2 extracted from the QPI measurements in region 1 (SI Fig. S23) and region 2 (SI Fig. S25) plotted in the continuum model calculated 31.1° band structures using the same parameters as in Fig. 4d but along a different k-path γ − m − γ, as marked in SM BZs in the inset. The error bars are shown as grey lines. Scale bars: a, b, c, 1nm.

## Methods

### Sample growth for STM

High-quality buffer on SiC was synthesized using a two-step process. First, the monolayer epitaxial graphene was synthesized using silicon sublimation from the Si face of the semi-insulating SiC substrates (II–VI). Before the growth, the SiC substrates were annealed in 10% hydrogen (balance argon) at 1,500 °C for 30 min to remove subsurface damages due to chemical and mechanical polishing. Then monolayer epitaxial graphene (MLEG) was formed at 1,800 °C for 30 min in a pure argon atmosphere. Second, a Ni stressor layer was used to exfoliate the top graphene layer to obtain fresh and high-quality buffer on SiC. After this, 270 nm of Ni was e-beam deposited on MLEG at a rate of 5 Å s$^{-1}$ as a stressor layer. Then a thermal release tape was used to peel off the top graphene layer from the substrate. The growth of $WSe_2$ crystals on an epitaxial graphene substrate was carried out at 800 °C in a custom-built vertical cold-wall chemical vapor deposition (CVD) reactor for 20 min[43]. The tungsten hexacarbonyl ($W(CO)_6$) (99.99%, Sigma-Aldrich) source was kept inside a stainless-steel bubbler in which the temperature and pressure of the bubbler were always held at 37 °C and 730 torr, respectively. Mass-flow controllers were used to supply $H_2$ carrier gas to the bubbler to transport the $W(CO)_6$ precursor into the CVD chamber. The flow rate of the $H_2$ gas through the bubbler was maintained at a constant 8 standard cubic centimeters per minute (sccm), which resulted in a $W(CO)_6$ flow rate of $9.0 \times 10^{-4}$ sccm at the outlet of the bubbler. $H_2Se$ (99.99%, Matheson) gas was supplied from a separate gas manifold and introduced at the inlet of the reactor at a constant flow rate of 30 sccm.

### Bulk $WSe_2$ growth

$WSe_2$ bulk crystals were grown using a flux synthesis method. 99.999% W powder and 99.999+% selenium shot were loaded into a quartz ampule at a 1:5 molar ratio and sealed under high vacuum ($\sim 10^{-6}$ Torr). These sealed ampules were then gradually heated to 1000 °C for over 48 h, followed by an eight-day dwell at 1000 °C. The ampules were subsequently cooled at a rate of 1.04 °C/h until they reached 750 °C at which point the ampule was removed from the furnace and quenched in water. After this step, the crystals and excess Se were transferred to a fresh ampule with quartz wool, and then sealed under vacuum. The ampule was heated to 285 °C, flipped, and centrifuged to separate the crystals from the excess Se. Finally, the $WSe_2$ crystals were transferred to a third vacuum-sealed ampule and placed under a temperature gradient for 16 hrs, with 285°C at the hot end and room temperature at the cold.

### STM and STS measurements

STM and STS measurements for the ~31.1° twisted bilayer $WSe_2$ sample were conducted at 4.3 K in the ultra-high vacuum chamber, with a base pressure of $2.0 \times 10^{-11}$ torr. STM and STS measurements for the ~31.5° twisted bilayer $WSe_2$ sample were conducted at 77 K in an Omicron LTSTM, with a base pressure of $1.0 \times 10^{-10}$ torr. The W tip was prepared by electrochemical etching and then cleaned by in situ electron-beam heating. STM dI/dV spectra were measured using a standard lock-in technique, for which the modulation frequency was 758 Hz. Two different modes of STS were simultaneously used: (1) the conventional constant-height STS and (2) the constant-current STS.

### Sample fabrication for 4D-STEM measurements

The free-standing TEM samples were prepared with tear-and-stack methods and Polypropylene carbon (PPC) based dry transfer technique[44]. All $WSe_2$ layers were mechanically exfoliated from high-quality crystals[45] by Polydimethylsiloxane (PDMS) on 285 nm $SiO_2$/Si substrates. A droplet of 15% PPC/Anisole solution was spin-coated on top of a PDMS stamp and was heated for 2 minutes at 150°C to dry. Under a temperature of around 40°C, we tore part of the $WSe_2$ monolayer and twisted the rest part with certain angles, and finally released it at 150°C. The PPC was annealed away under 250 °C and untra high vacuum (UHV) ($1 \times 10^{-9}$ torr) for 72 hours.

The TEM samples with hBN cappings were prepared with tear-and-stack methods and Polycarbonate (PC) based dry transfer technique. 15% PC solution is made into PC thin films that cover onto a PDMS dome. A thin hBN flake (<15 nm) with a size larger than 25 um x 25 um was picked up by PC at 100°C. The WSe$_2$ monolayers exfoliated on Si wafers were firstly be torn into two pieces and then picked up respectively by the PDMS / PC / hBN stamp at 90°C with a relative twist angle controlled by a step-motor driven rotation stage (Optosigma OSMS60-YAW, accuracy 0.0025°). The PC / hBN / tWSe$_2$ was then dropped off onto a TEM grid (Norcada NH005D1) at 180°C. The PC residue can be washed off by Chloroform (30 mins) and IPA (10 mins). See the optical microscopy figure of our STEM sample in Fig. S27a-b.

**Sample fabrication for STM measurements**

The STM sample was prepared with tear-and-stack methods and both PC- and PPC-based dry transfer techniques. A few-layer-graphite was first picked up by the PDMS / PC stamp at 120°C. Then a thick hBN flake (~30 nm) with a size smaller than graphite was picked up by PDMS / PC / graphite stamp at 90°C. The resulting PC / graphite / hBN was dropped off at 180°C onto the 285 nm SiO$_2$ / Si wafer with Pt / Cr landing patterns made through the standard photolithography process. The PC can be washed off by Chloroform (30 mins, magnetic stirring on), Dichloromethane (30 mins, magnetic stirring on), Chloroform (30 mins, magnetic stirring on), and IPA (10 mins). The Graphite / hBN / SiO$_2$ / Si was then annealed at 400°C under 10$^{-6}$ Torr vacuum for 5 hrs to clean the possible PC residues on the surfaces. Atomic force microscopy (AFM) contact mode cleaning was then applied to the graphite surface to further remove the residues.

Afterward, we transferred the tWSe$_2$ onto graphite contact. The WSe$_2$ monolayers exfoliated on Si wafers were firstly torn into two pieces and then picked up respectively by the PDMS / PPC stamp at 40°C with a relative twist angle controlled by a step-motor driven rotation stage (Optosigma OSMS60-YAW, accuracy 0.0025°). Then, the PPC / tWSe$_2$ was dropped off onto graphite at 150°C. The PPC can be washed off by Acetone (30 mins) and IPA (10 mins). Afterward, the tWSe2 / Gr / hBN / SiO$_2$ / Si sample was annealed in an ultra-high vacuum (10$^{-10}$ Torr) at 250°C for 24 hrs to reduce the PPC residues. AFM contact mode cleaning was then applied to the tWSe$_2$ surface to further remove the polymer residues. Before being measured by the STM, another UHV annealing (10$^{-10}$ Torr) at 250°C in the STM chamber was performed for 24 hrs. After all these pretreatments, the atomic clean tWSe$_2$ STM sample was achieved. See the optical microscopy figure of our STM sample in Fig. S27c.

**ADF-STEM data acquisition**

ADF-STEM imaging (Fig. 1c, e, f, 2d) was conducted using an aberration-corrected ThermoFisher Titan Themis operated at 80 kV with an approximately 15 pA probe current. The acquisition time per pixel was 200 ns for each frame and multiple frames (20-50) were acquired and cross-correlated afterward to improve the signal-to-noise ratio and reduce the scan noise introduced by the sample drift. A 30 mrad convergence angle and an 80-200 mrad collection angle were used for all ADF-STEM images, whose contrast is proportional to $Z^\gamma$, where Z is the atomic number and $1.3 < \gamma < 2$.

**4D-STEM data acquisition and processing**

The 4D-STEM datasets were collected using an aberration-corrected ThermoFisher Titan Themis equipped with an Electron Microscope Pixel Array Detector (EMPAD)[46] at an acceleration voltage of 80 keV. The EMPAD features a 128 × 128 pixel array, a 1 ms acquisition time, a readout speed of 0.86 ms per frame, and a dynamic range of 1,000,000:1. Two distinct types of 4D-STEM datasets were acquired: large convergent beam 4D-STEM data and nano-beam 4D-STEM data.

For the large convergent beam 4D-STEM datasets, a convergence angle of 30 mrad and a defocus value of approximately -20 nm were employed to reconstruct high-resolution atomic images of moiré lattices (Fig. 2b). The scan step size was set to 1.168 Å, and the electron dose rate was approximately $\sim 10^4$ e$^-$ Å$^{-2}$. The mixed-state electron ptychography reconstruction method[47] asssisted by large language model[48] was utilized on these datasets with a 12 probe mode, achieving a reconstruction resolution of approximately 0.6 Å.

Nano-beam 4D-STEM datasets (Fig. S2, S3), utilizing a 0.5 mrad convergence angle, were employed to ascertain lattice information across the entire sample. A camera length of 720 mm was used to ensure a clear delineation of the twelve first-order diffraction spots. Analysis of the center of mass of these spots facilitated the mapping of lattice strain, layer rotation, and the twist angle between two layers with sub-picometer precision[49] in a large field of view (approximately 600 nm). This expansive view encompasses the entire region of interest in the bilayer samples.

**Theoretical calculation**

The density functional theory calculations were implemented in the Quantum ESPRESSO suite[50,51]. We used the Perdew–Burke–Ernzerhof exchange-correlation functional[52] in all density functional theory calculations. The structural optimization was obtained with the criteria for force less than 0.025 eV Å$^{-1}$, pressure less than 0.5 kbar, and total energy less than 0.0014 eV. We used optimized norm-conserving pseudopotentials[53] from the PseudoDojo library[54] with plane waves kinetic energy cutoff of 98 Ry as recommended. The vdW interaction was considered within the semiempirical approach Grimme-D3[55]. The spin–orbit coupling was also included. The unfolded band structures were performed using the BandUP code[56–58].

**Method References**

**Acknowledgement**

We acknowledge Prof. Young-Woo Son for the useful discussion. Y.L., F.Z., H.K. and C.-K.S. are supported by the NSF through the Center for Dynamics and Control of Materials: an NSF Materials Research Science and Engineering Center under cooperative agreement no. DMR-2308817, the US Air Force grant no. FA2386-21-1-4061, NSF grant nos. DMR-1808751 and DMR-2219610, and the Welch Foundation F-2164. C.S. and Y.H. acknowledge the support from NSF grant no. CMMI-2239545 and the Welch Foundation C-2065. V.-A.H. and F.G. were supported by the Welch Foundation under Grant No. F-2139-20230405. X. Liu and X. Li were partially supported by the Department of Energy, Office of Basic Energy Sciences under grant DE-SC0019398 for device fabrication and the Welch Foundation chair F-0014 for sample preparation. C. D. and J. A. R. were supported by the Penn State Center for Nanoscale Science (NSF grant no. DMR-2011839) and the Penn State 2DCC-MIP (NSF grant no. DMR-1539916). Y.-C.L. acknowledges the support from the Center for Emergent Functional Matter Science (CEFMS) of NYCU and the Yushan Young Scholar Program from the Ministry of Education of Taiwan. L.N.H, M.H, J.H, and K.B are supported by NSF MRSEC program at Columbia through the Center for Precision-Assembled Quantum Materials, NSF grant no. DMR-2011738. Y.X., Q.G., and E.K. are supported by the NSF through the Center for Dynamics and Control of Materials: an NSF Materials Research Science and Engineering Center under cooperative agreement no. DMR-2308817.

**Author Contributions**

C.-K.S. and Y.L. conceived the experiment. Y.L. and F.Z. carried out the STM and STS measurements. C.S. and Y.H. carried out the STEM measurements. C.S. and Y.J. performed electron ptychography reconstruction. Y.X., Q.G., and E.K. performed the theoretical model calculations. N.M-D. was involved in the discussion of the continuum model of TMDs.  V.-A.H. and F.G. performed the DFT calculations. Y.-C.L. synthesized the twisted WSe$_2$ bilayers. C.D. prepared the graphitic buffer layer/SiC. J.A.R. supervised the sample preparation effort. H.K. annealed and pre-treated the sample. X. Liu and Y.L. prepared the exfoliated the samples for STEM and STM under the supervision of X.Li. L.N.H, M.H, K.B. and J.H. synthesized the WSe$_2$ bulk crystals. Y.L., F.Z., Y.H., C.S. and C.-K.S. analyzed the STM and STEM data. Y.L., C.S., Y.H., E.K and C.-K.S. wrote the paper with contributions from all the authors. † These authors contributed equally to this work.

**Competing Interests:** All the authors declare no competing interests.

**Data Availability:** Source data that reproduces the plots in the main text and extended data figures are provided with this paper. Source data that reproduce the plots in supplementary information are available upon request.

**Code Availability:** Source codes used in the manuscript are provided with the paper. The DFT calculations presented in the paper were carried out using publicly available electronic structure codes (referenced in Methods). All other codes in supplementary information are available upon reasonable request.

**Extended Figure Captions**

**Extended Figure 1 | STM/S measurement of a fabricated ~31.5° twisted bilayer WSe$_2$ SM. a,** An STM image of the 31.5° twisted bilayer WSe$_2$, showing a <$10^{11}$ cm$^{-2}$ low density of defects thanks to the high-quality bulk crystal WSe$_2$ ($V_{Bias}$ = -0.98 V, I = 100 pA). **b,** An in-gap taken STM topography image of ~31.5° tWSe$_2$ ($V_{Bias}$ = -0.6 V, I = 100 pA), revealing unambiguous SM patterns. **c,** The FFT image of b. The black squares, blue circles, and red diamonds represent top layer WSe$_2$ Bragg peaks, second order and first order tWSe$_2$ moiré, with SM splitting, respectively. **d,** Filtered iFFT image using the FFT peaks in red diamonds. **e,** The logarithmic scale constant height STS taken in the pristine region ($V_{set}$ = -2.0 V, I = 100 pA, $V_{amp}$ = 29 mV). Blue and green arrows represent the $\Gamma_1$ and $K_{VBM}$ respectively. **f,** The constant current STS taken at the same site ($V_{int}$ = -1.4 V, I = 50 pA, $V_{amp}$ = 19 mV). Blue and green arrows represent the $\Gamma_1$ and $K_{VBM}$ respectively. Scale bars: 3 nm.

**Extended Figure 2 | QPI measurement and analysis on the fabricated ~31.5° twisted bilayer WSe$_2$ SM. a-c,** The constant current dI/dV mappings taken at -1.0 V (a), -0.98 V (b), and -0.96 V (c) (I = 100 pA, $V_{amp}$ = 20 mV). **d,** Schematic of the intra-K-valley scattering in ~31.5° SM. The purple and grey BZs depict the CM and SM BZs, respectively. $\Delta_{SM}$ is the SM minigap while $\Delta_{CM}$ is the CM minigap. The blue arrow exhibits the q vector from intra-K-valley scatterings. The SM γ overlaps with CM $\Gamma_m$ and $K_m$ valleys. **e-g,** The FFT images of a-c, Red boxes label the regions of QPI q vector rings near the center. Insets: magnified near-zero FFT frequency images in the red boxes, showing the rings of QPI. **h,** E v.s. q/2 plot extracted from the QPI of ~31.5° tWSe2 sample. The QPI measured under 77K can infer the intra-K-valley scatterings information. The blue curve shows the parabolic fitting of K-valley band dispersion from the three QPI measurement dots. The fitted effective mass and $K_{VBM}$ energy levels are $m_{eff}$ = 0.55± 0.1, $E_{KVBM}$ = -0.925 ± 0.1 eV. The error bars are shown as grey lines.

Figure 1

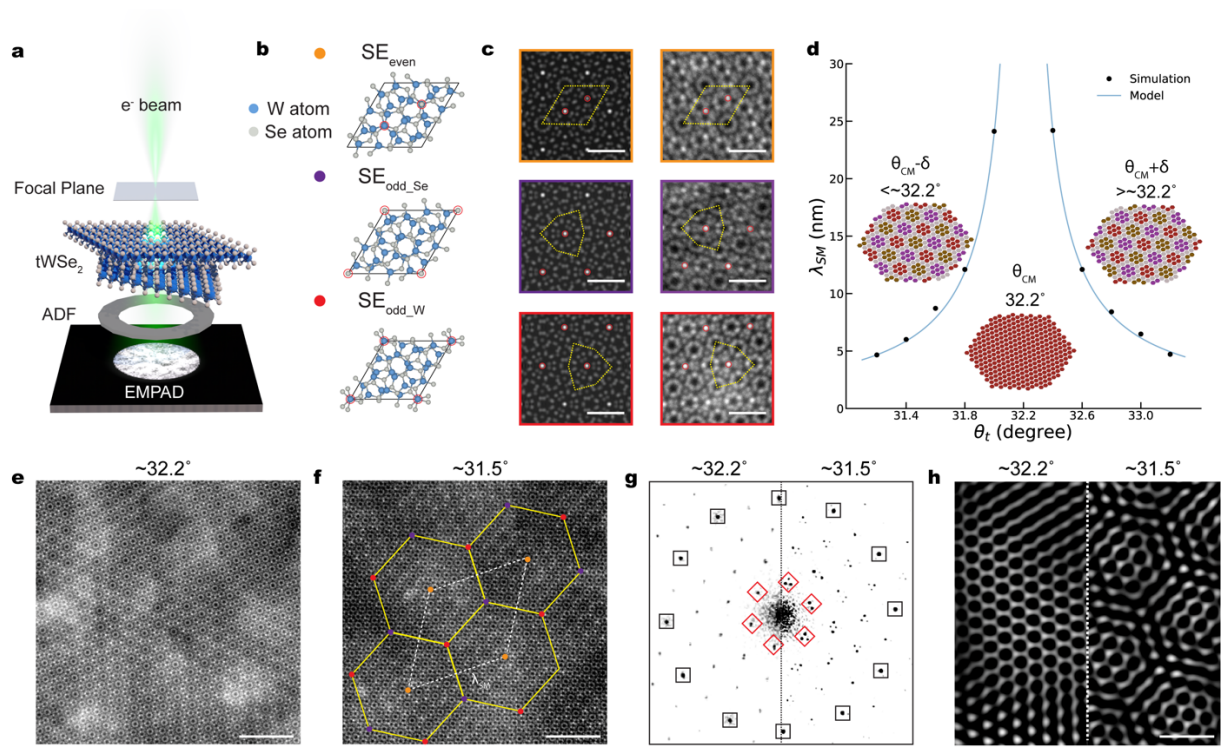

Figure 2

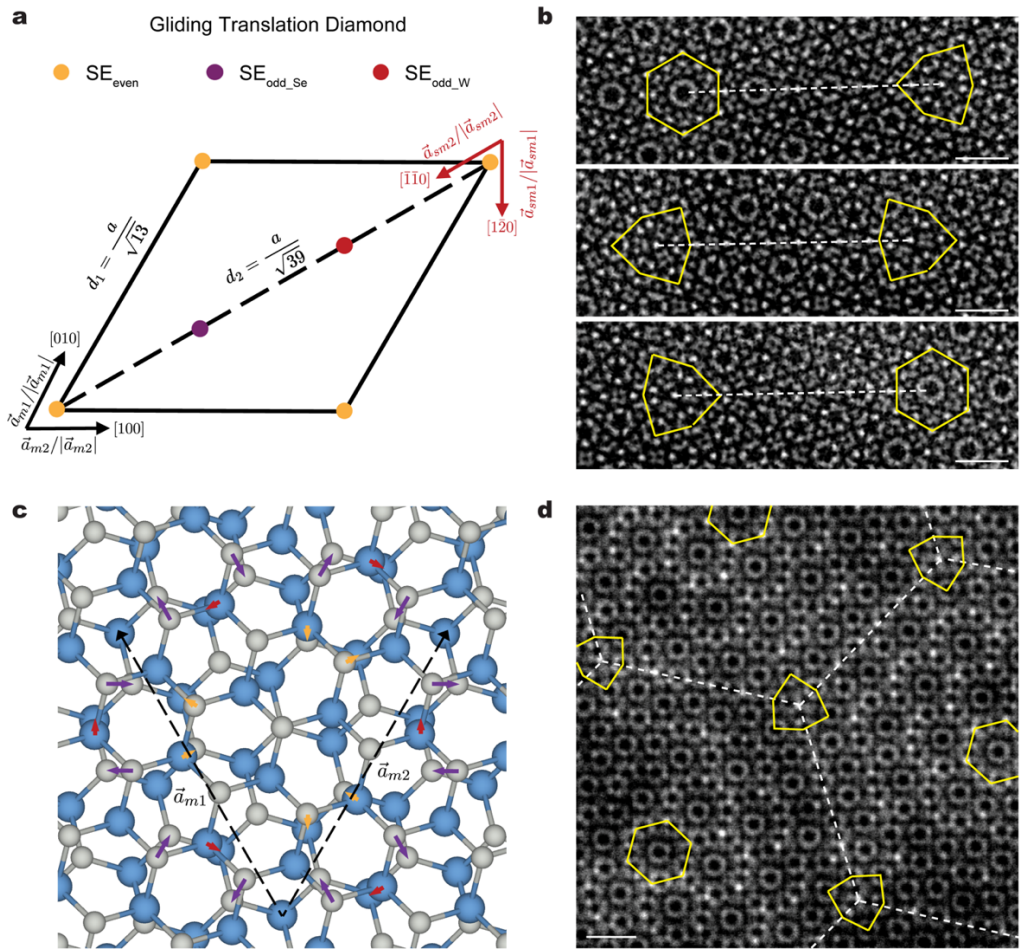

Figure 3

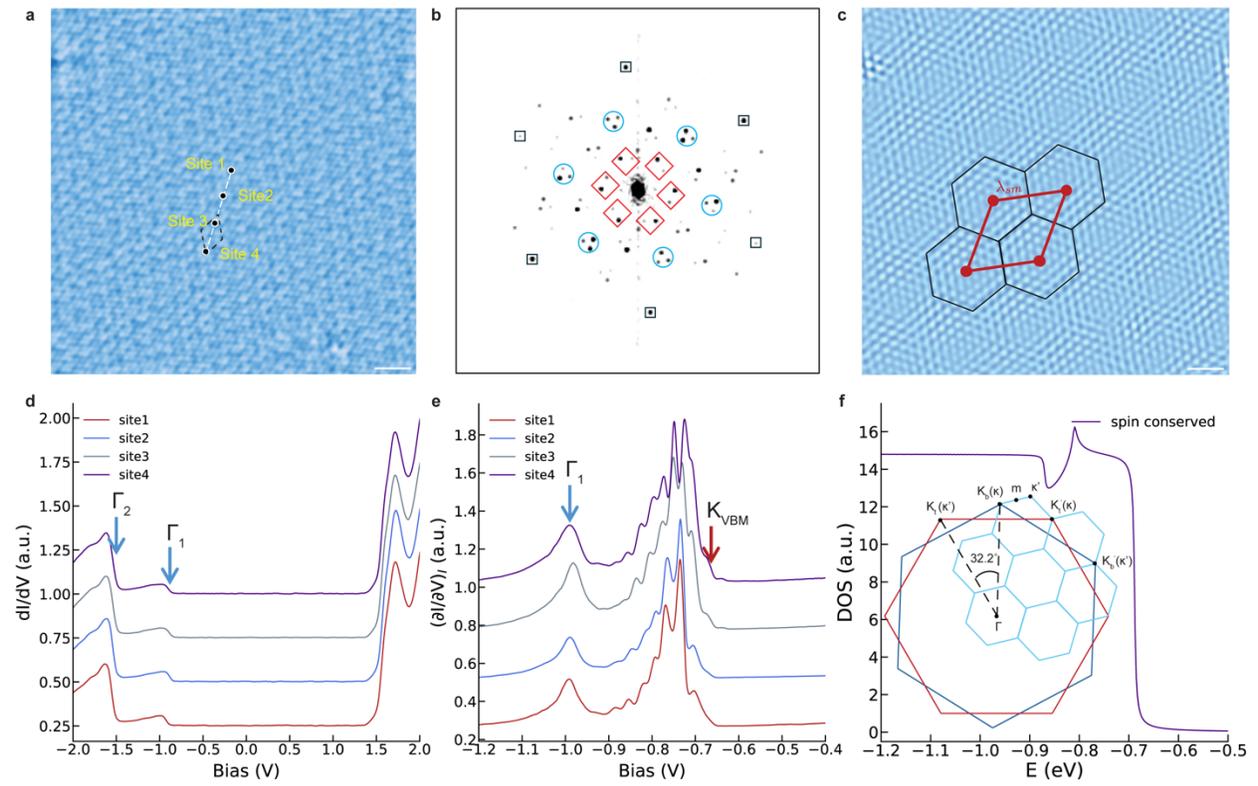

Figure 4

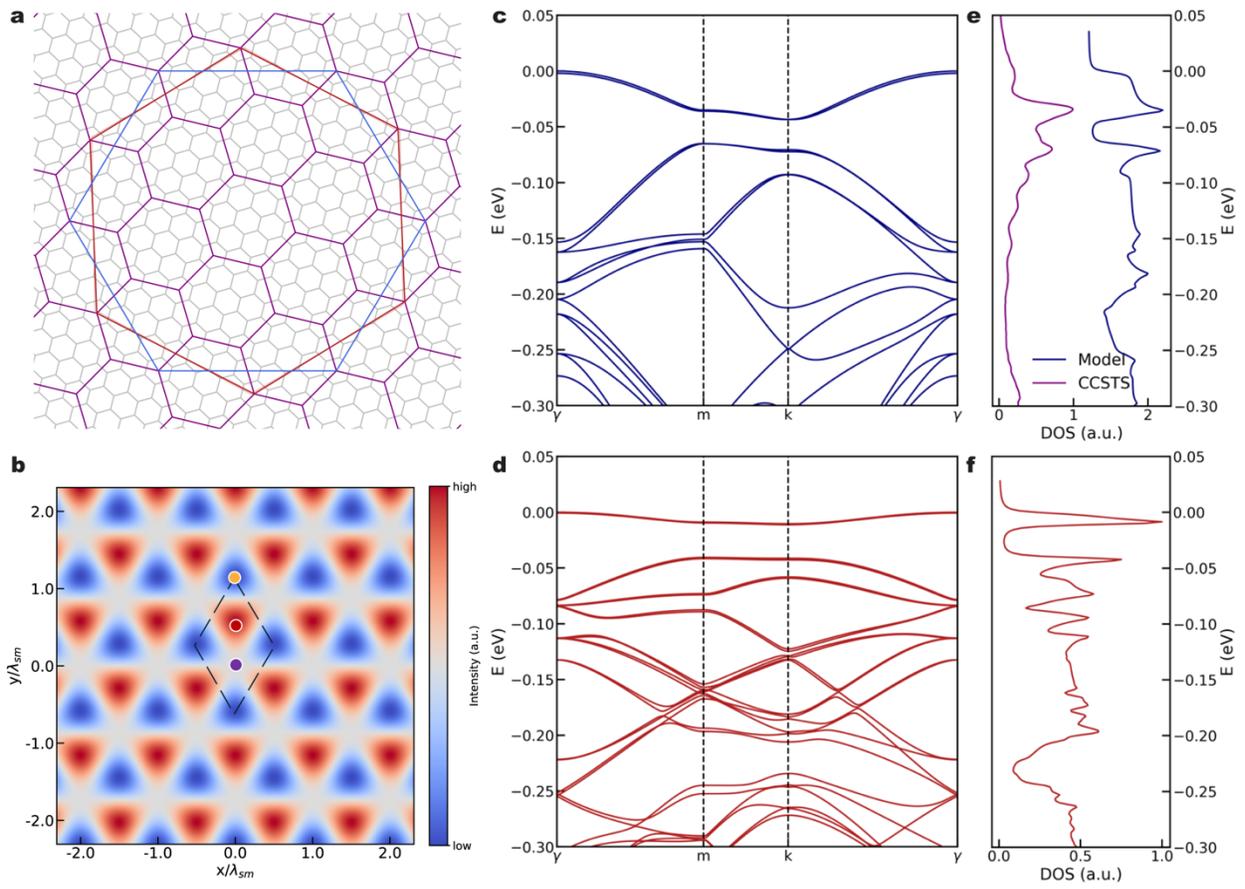

Figure 5

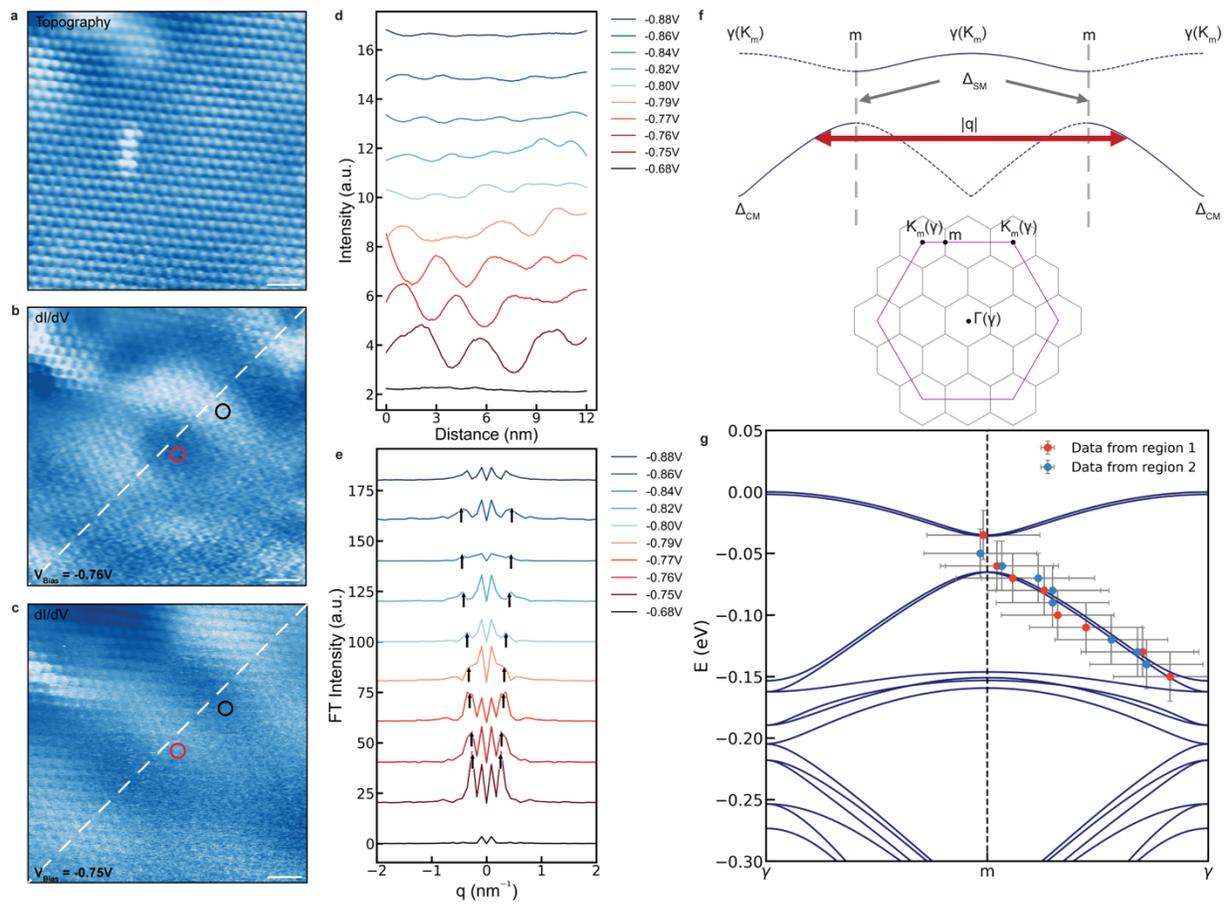

Extended Figure 1

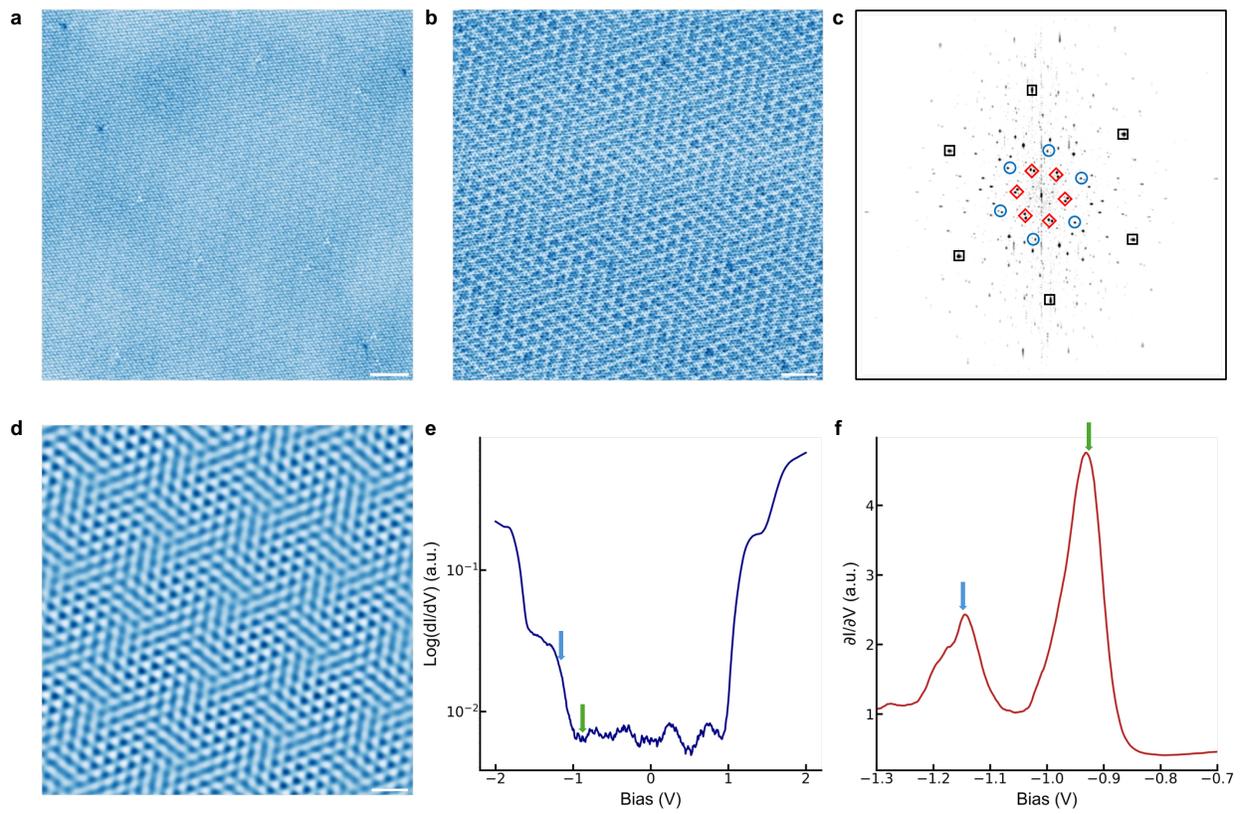

# Extended Figure 2

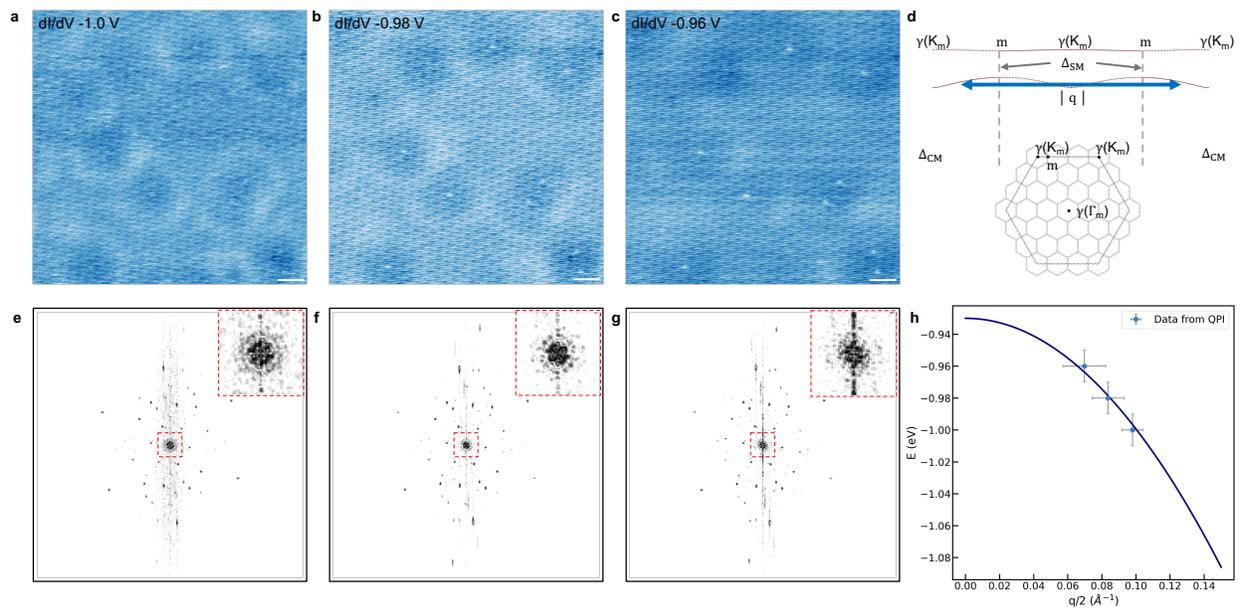